\documentclass[10pt, final, conference, compsocconf]{IEEEtran}

\usepackage{wrapfig}
\usepackage{amsmath,amssymb,amsthm,supertabular,booktabs,graphicx,subfigure,multirow,semantic}
\usepackage[all]{xy}
\usepackage{microtype}
\usepackage{cite}
\usepackage{fixltx2e}

\author{
\IEEEauthorblockN{Ivan J. Jureta}
\IEEEauthorblockA{PReCISE Research Center, University of Namur\\ ivan.jureta@fundp.ac.be}
}

\begin{document}

\theoremstyle{plain}
\newtheorem{thm}{Theorem}[section]
\newtheorem{proposition}[thm]{Proposition}
\newtheorem{corollary}[thm]{Corollary}
\newtheorem{lemma}[thm]{Lemma}
\newtheorem{definition}[thm]{Definition}
\newtheorem{convention}[thm]{Convention}

\theoremstyle{remark}
\newtheorem{remark}[thm]{Remark}
\newtheorem{example}[thm]{Example}

\newcommand{\zi}[1]{\textit{#1}}
\newcommand{\xb}[1]{\textbf{#1}}
\newcommand{\xf}[1]{\textsf{#1}}
\newcommand{\xc}[1]{\textsc{#1}}
\newcommand{\xt}[1]{\texttt{#1}}

\newcommand{\eendex}{$\blacksquare$}
\newenvironment{xexample}[1]{\begin{example}\label{#1}}{\eendex\end{example}}

\title{Requirements Engineering Methods:\\A Classification Framework and Research Challenges}

\maketitle

\thispagestyle{empty}

\begin{abstract}
Requirements Engineering Methods (REMs) support Requirements Engineering (RE) tasks, from elicitation, through modeling and analysis, to validation and evolution of requirements. Despite the growing interest to design, validate and teach REMs, it remains unclear what components REMs should have. A classification framework for REMs is proposed. It distinguishes REMs based on the domain-independent properties of their components. The classification framework is intended to facilitate (i) analysis, teaching and extension of existing REMs, (ii) engineering and validation of new REMs, and (iii) identifying research challenges in REM design. The framework should help clarify further the relations between REM and other concepts of interest in and to RE, including Requirements Problem and Solution, Requirements Modeling Language, and Formal Method.
\end{abstract}


\section{Introduction}\label{s:introduction}
A Requirement Engineering Method (\xc{rem}) can be thought of as a combination of a formalism for the representation and analysis of requirements, and of processes to support and guide the user of the formalism through Requirements Engineering (\xc{re}) tasks, such as requirements elicitation, representation, validation, verification, and evolution. Examples of \xc{rem}s are \xc{rml} \cite{Greenspan:1984:PHD}, \xc{erae} \cite{Dubois+:1988:PJR}, \xc{nfr} \cite{Mylopoulos+:1992:TSE}, \xc{kaos} \cite{Dardenne+:1993:SCP}, i* \cite{Yu+:1994:ICSE}, Viewpoints \cite{Finkelstein+:1994:TSE}, Labeled Quasi-Classical Logic \cite{Hunter+:1998:TOSEM}, Tropos \cite{Castro+:2002:IS}, Formal Tropos \cite{Fuxman+:2004:REJ}, \xc{carl} \cite{Gervasi+:2005:TOSEM}, Techne \cite{Jureta+:2010:RE}, and many others. 

\begin{figure}[b!]
\vspace{-8mm}
	\centering
	\includegraphics[width=7.5cm]{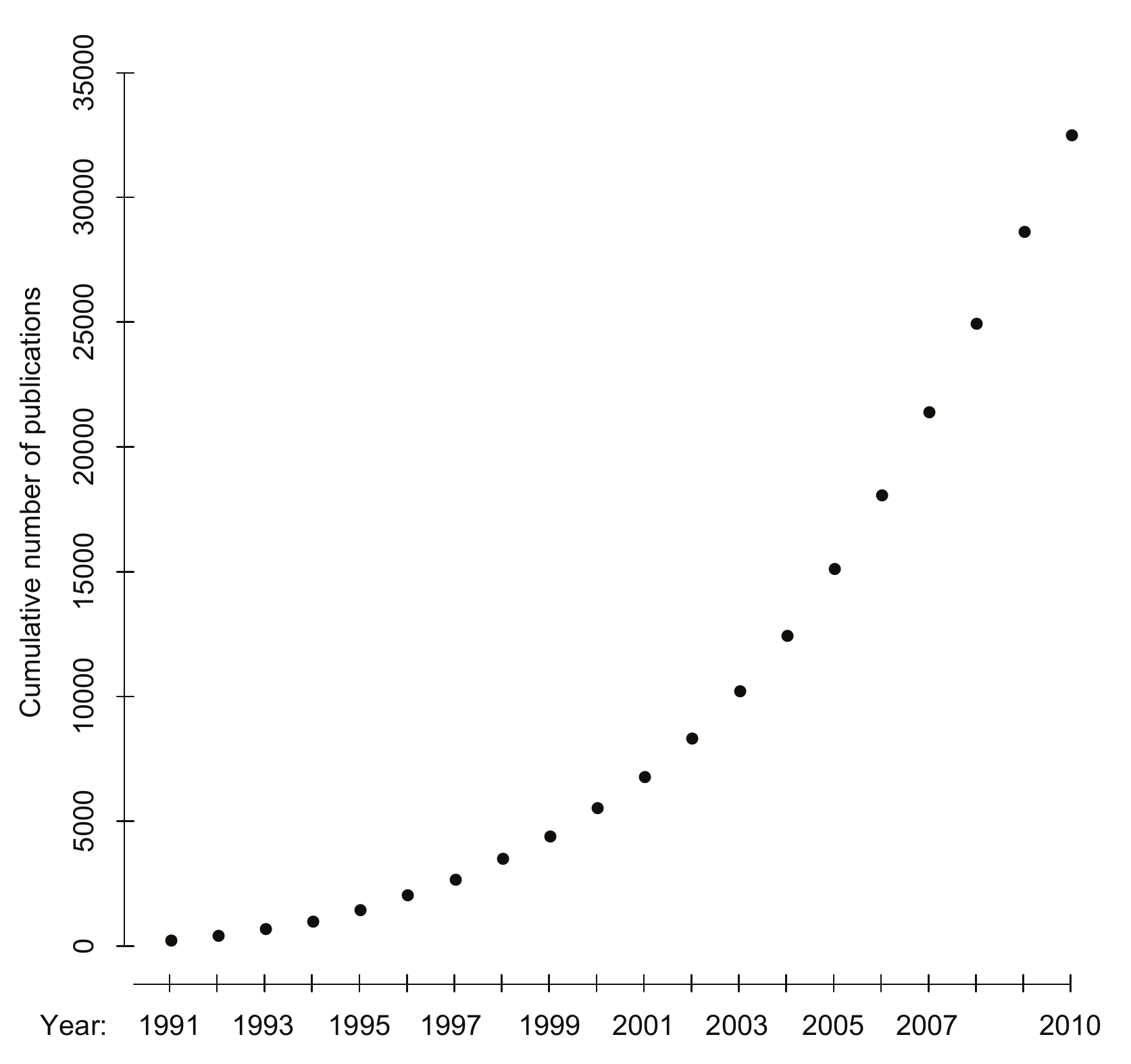}
\vspace{-3mm}
\caption{Cumulative number of publications since 1991 that cite in the title, abstract or main text the exact phrase ``requirements engineering'' together with one or more of the terms ``method'', ``approach'', ``framework'', ``methodology''. Data from Google Scholar.}
\label{f:cumulative-publications}
\end{figure}

It is difficult to estimate the number of \xc{rem}s and of the publications on \xc{rem}s. \zi{Method} informally means a procedure for doing something, so really \zi{any} publication which proposes \zi{how to do something} within the scope of \xc{re} proposes \zi{a method for} \xc{re}, regardless of what that method's intended scope (coverage) and depth (level of detail) may be. Figure \ref{f:cumulative-publications} gives a very rough estimate of the cumulative number of publications related to \xc{rem}s since 1991. \xc{rem}s are also of interest outside \xc{re}, such as in Business Analysis \cite{HP:2007,BABOK:2009}. 

Despite the interest in \xc{rem}s, it is still unclear how to answer basic questions about them. Which components does an \xc{rem} have? Which components must it have and why? When is an \xc{rem} domain-specific? Given two \xc{rem}s, how can we compare them? How to systematically make an \xc{rem}? How to know if some contribution in \xc{re} is an \xc{rem}, an \xc{rem} component, or something else? How does \xc{rem} research relate to formal methods, logics, ontology engineering? What to include in a course on \xc{rem} design?


This paper does not answer all of these questions. Instead, the proposal is a classification framework for \xc{rem}s that distinguishes \xc{rem}s based on the properties of their components. A component is a set of concepts, relations, rules, or other tools having a well-delimited role within \xc{rem}. The framework uses five components, Requirements Problem and Solution, Ontology, Formalism, Organization Mechanism, and Guidelines. The classification framework is intended to help (i) the analysis, teaching and extension of existing \xc{rem}s, (ii) the engineering and validation of new \xc{rem}s, and (iii) the identification and organization of research challenges in \xc{rem}s design and validation. Classification dimensions other than components are certainly relevant, but stay outside the scope of this paper (e.g., classification by scope, by domain, by results from use, etc.).

The framework is first introduced and illustrated (\S\ref{s:components}), research challenges (\S\ref{s:challenges}) and limitations (\S\ref{s:discussion}) are discussed, and conclusions are summarized (\S\ref{s:conclusions}).

\section{Classification Framework}\label{s:components}
The classification framework uses specific terminology, capitalized hereafter and italicized when introduced first.

A \zi{Component} in the classification framework is a set of concepts, relations, rules, or otherwise, which together serve a specific purpose within an \xc{rem}, such as, e.g., to categorize requirements, to visualize requirements, etc. Each Component comes with \zi{Component Properties}. Component Properties are \zi{domain-independent}, which in this paper means that a Component Property is independent from a ``paradigm'' followed in the design of an \xc{rem}, so that there is nothing in the framework to make it specific to goal-oriented \xc{rem}s, or others.

The classification framework uses five Components:

\begin{enumerate}
	\item{\xb{Requirements Problem and Solution} that the \xc{rem} should help, respectively, define and find;}
	\item{\xb{Ontology}, defining categories of information input to, used and output by an \xc{rem};}
	\item{\xb{Formalism}, for representation and analysis instances of the concepts and relations in the Ontology;}
	\item{\xb{Organization Mechanism} for the organization of representations made with the Formalism;}
	\item{\xb{Guidelines}, advising how to use the Ontology, Formalism and Organization Mechanisms to define the Requirements Problem and find Solutions to it.}
\end{enumerate}

The overall idea behind the framework is simple. The purpose of an \xc{rem} needs to be explicit, and is conveyed through the Requirements Problem that it should help solve, and the Solution it should help produce. The Ontology identifies the information that the \xc{rem} will manipulate. That information ought to be represented in some structured way, to help answer questions about requirements. Because there can be a considerable amount of information to manipulate, there are Organization Mechanisms, to help decompose and manage representations. Finally, Guidelines will say why and how to use the conceptual tools, namely the Ontology, Formalism, and Organization Mechanisms to instantiate the Requirements Problem to a specific system-to-be, and find and describe Solution instances.

The rest of this section discusses Components and Component Properties. Each Component is presented in the same way, with (i) a definition of the Component, (ii) the purpose of the Component in an \xc{rem}, (iii) Component Properties, (iv) examples that illustrate Component Properties in existing \xc{rem}s, and (v) discussion of Component Properties in relation to the hypothetical \xc{rem} in the case study.

\subsection*{Case Study}
Suppose that the aim is to design an \xc{rem}, call it X, that should help solve the requirements problem as Zave \& Jackson \cite{Zave+:1997:TOSEM} defined it. The problem says that, given a set of requirements that must be satisfied together (denote the set $R$) and domain assumptions which must be satisfied together ($K$), we should find a specification of properties and behaviors of the system-to-be ($S$) such that $K, S \vdash R$, where $\vdash$ is the consequence relation of classical logic.


\subsection{Requirements Problem and Solution}\label{s:components:requirements-problem-and-solution}

\subsubsection{Definition} The Requirements Problem concept defines the undesirable properties of the situation at the start of \xc{re}, which the \xc{rem} should help solve. The Solution concept defines desirable properties of the result that the requirements engineer aims to make with the \xc{rem}.

\subsubsection{Purpose of Req. Problem and Solution in \xc{rem}}
\begin{itemize}
\item{Define the purpose of \xc{rem}.}
\item{Define the desired result of applying the \xc{rem}.}
\item{Force the \xc{rem} designer to clearly state what her \xc{rem} is intended to do within \xc{re}, thereby forcing her to state the scope and depth at which she views the problem that the \xc{rem} should help solve.}
\item{Compare \xc{rem}s, in that the Problem and Solution statements help us evaluate if an \xc{rem} is focusing on the same problem as another, if it focuses on a more specific, or a more general problem.}
\item{Justify design choices in the \xc{rem}: Decisions to include some concepts and relations in the Ontology, support specific rules of reasoning in the Formalism, and include processes in the Guidelines can be justified through their relevance to the description of the Requirements Problem, and finding and description of Solution instances.}
\end{itemize}

\subsubsection{Component Properties}
\begin{itemize}
\item{\zi{Implicit Definition}: No explicit statement of the Requirements Problem and Solution concepts is given, but can be inferred from other Components of the \xc{rem}.}
\item{\zi{Informal Definition}: Component Property is satisfied if the Requirements Problem and Solution concepts are defined in natural language.}
\item{\zi{Formal Definition}: The Requirements Problem and Solution are defined as expressions in a Formalism.}
\end{itemize}

\subsubsection{Examples}
As there are no explicit definitions of Requirements Problem and Solution statements, all \xc{rem}s cited up to this point, except for Techne have Implicit Definitions. For example, it is clear in Tropos and \xc{kaos} that the aim is to find operationalizations of high-level goals, but the explicit statement of this problem, such as the one given by Zave \& Jackson, is absent. Techne is an example where both Informal and Formal Definition are used. 

\subsubsection{Case Study}
X is intended to solve the Zave \& Jackson statement of the Requirements Problem, and this problem has both an Informal and a Formal Definition, given in the Case Study. The Solution concept is implicit in the definition of the Requirements Problem, as it is a specification $S$ which is consistent with $K$ and $R$ (otherwise it cannot be that $K, S \vdash R$), and which together with domain assumptions is enough to derive requirements, i.e., $K, S \vdash R$.

\subsection{Ontology}\label{s:components:ontology}

\subsubsection{Definition} An Ontology in an \xc{rem} is an explicit specification of concepts and relations, whose instances are input, used and output by the \xc{rem}.

\subsubsection{Purposes of an Ontology in \xc{rem}}
\begin{itemize}
\item{Scoping: By including some and excluding other concepts and relations, the Ontology identifies the categories of information judged relevant to achieve the purpose of the \xc{rem}.}
\item{Communication: Concepts and relations of the Ontology give the starting stable set of terms to use in communication about requirements.}
\item{Documentation: Categories of information in the Ontology should be captured by artifacts that document requirements. Ontology helps structure these artifacts.}
\item{Focusing: For the engineer/user of the \xc{rem}, the Ontology acts as a checklist information to focus on when applying the \xc{rem}.}
\end{itemize}

\subsubsection{Component Properties}
\begin{itemize}
\item{\zi{Implicit Definition}: Concepts and relations are not explicitly defined, but can be inferred from other Components of the \xc{rem}. In such cases, it appears as if there is no Ontology; it can, however, be determined by looking at the kinds of information used by other Components.}
\item{\zi{Informal Definition}: Informal interpretation for concepts and relations is given via definitions in natural language, the choice of concepts and relations is justified with regards to the purpose of the \xc{rem}, and a discussion given of the ontological commitments, i.e., the assumptions for choosing particular concepts and relations rather than others, and for defining them exactly as proposed.}
\item{\zi{Structured Definition}: Concepts and relations are represented as a graph, where concepts are nodes and relations are edges of the graph. This is the case when, e.g., an Entity-Relationship model is used to describe the Ontology of the \xc{rem}.}
\item{\zi{Formal Definition}: Concepts and relations are defined using expressions of a formal logic.}
\end{itemize}

\subsubsection{Examples}
In \xc{kaos}, the conceptual meta-model provides the Structured Definition of the Ontology. Informal definitions are given in intensional form, listing properties of concepts (e.g., properties of the goal concept) and properties of relations (e.g., minimality and consistency for the goal refinement relation). Similar combination is applied in Tropos, Formal Tropos, i*. To the best of my knowledge, there are no \xc{rem}s with Formal Definitions for their complete Ontology. The Core Ontology for Requirements, subsequently used in Techne, accomplishes this only in part and indirectly, by mapping its concepts to a foundational ontology which has a Formal Definition (see Appendix B of \cite{Jureta+:2009:AO}). For illustration of Formal Definition of Ontology, but unrelated to \xc{re}, see \xc{dolce} \cite{Masolo+:2003}. \xc{carl} is a case of Informal Definition. Implicit Definition occurs in the original presentation of both Viewpoints and Labeled Quasi Classical Logic. In both cases, concepts and relations are implicit in the Formalisms and Guidelines, but explicit and separate definitions are not given.

\subsubsection{Case Study}
The Ontology for X should include the Requirement, Domain assumption, and Specification concepts. Assume for simplicity that all are top-level concepts. The consequence relation $\vdash$, since it is from classical logic, tells us the Ontology should include a Satisfaction relation from the Domain assumption and Specification concepts to the Requirements concept. Note that $\vdash$ is about derivability, not satisfaction, but classical logic is sound and complete so we can talk about Satisfaction as it fits the intuition that requirements are there to be satisfied. 

Informal Definition in X consists of giving (at least) natural language definitions for the K, S, R concepts and all relations, then justifying why there are the three concepts and not more or less, and why they are top-level (i.e., one is not a specialization of another). An Informal Definition of the Requirement concept is that it is an optative statement about the environment of, and/or about the system-to-be. A Structured Definition for X can be an Entity-Relationship diagram, showing the three concepts as nodes, and relations as links. A Formal Definition can consist of defining the concepts using predicates of a more abstract ontology, such as writing that $\zi{Optative}(\phi) \rightarrow \zi{Requirement}(\phi)$, to say that $\phi$ is a requirement if it is optative, whereby the predicate $\zi{Optative}$ would be defined in the more abstract ontology. 



\subsection{Formalism}\label{s:components:formalism}

\subsubsection{Definition} A Formalism serves for the representation of, and reasoning about instances of concepts and relations of the Ontology. A Formalism in \xc{rem} defines (i) symbols, (ii) rules for combining symbols into expressions, whereby the expressions refer to instances of the concepts and relations of the Ontology, (iii) semantic domain and semantic mapping function, to assign values within a domain of interest to expressions, and (iv) rules and algorithms for making deductions from, and/or checking properties of, expressions. 

\subsubsection{Purposes of a Formalism in \xc{rem}}
\begin{itemize}
\item{Representation/Modeling: Expressions written in the Formalism are models of information elicited for, and used in \xc{rem}. Modeling helps reflection on requirements and helps highlight the relationships between requirements.}
\item{Communication and Learning: Models facilitate communication between stakeholders and help new project participants to learn about the requirements of the system-to-be.}
\item{Analysis: If the Formalism has the necessary features, models can automatically be checked for properties of interest, such as consistency, completeness, presence of solutions to the requirements problem that the model defines.}
\item{Prediction: If the Formalism has the necessary features, simulations can be performed on models, to evaluate, e.g., the probability of failure of a requirement, given a particular way to operationalize it.}
\item{Traceability: Provided that the Formalism allows the distinction of model versions and for capturing the rationale for version changes, the Formalism can help document traces and aid traceability.}
\end{itemize}

\subsubsection{Component Properties}
\begin{itemize}
\item{\zi{Multi-Formalism}: \xc{rem} uses more than one Formalism.}
\item{\zi{Syntax Properties}: Properties of alphabet and of grammar (i.e., rules for combining symbols in the \xc{rem}):
\begin{itemize}
\item{\zi{Symbolic Syntax}: Models are well formed formulas as in a formal logic; it is relevant then to look into properties such as the presence of labels on formulas (to indicate sorts, or to keep track of formulas in deductions, as in labeled deduction), of predicates, quantifiers, and so on.}
\item{\zi{Graphical Syntax}: Models are drawn as diagrams; it is relevant then to look into the properties of graphs that these diagrams define, how the diagrams evaluate on cognitive effectiveness criteria \cite{Moody+:2010:REJ}, etc.}
\item{\zi{Syntax Maps}: Presence of rules to map expressions written in one syntax to expressions written in another syntax, in order to indicate that the expressions refer to the same (or aspects of the same) instances of concepts and relation (i.e., that the expressions aim to state the same information). This Component Property applies to \xc{rem}s which include two or more ways to represent the same information (e.g., symbolic and graphical syntax)}
\end{itemize}}
\item{\zi{Deductive System Properties}: Properties of the set of rules capturing correct inferences from a given set of expressions; properties of interest include:
\begin{itemize}
\item{\zi{Classicality}: How the Deductive System in the \xc{rem} relates to that of classical logic; this can be established by verifying which of Gabbay's \cite{Gabbay:1985} 13 properties the \xc{rem}'s Deductive System satisfies.}
\item{\zi{Paraconsistency}: Whether the Deduction System allows deriving any formula from an inconsistent set of formulas; property relevant for inconsistency handling, as a Paraconsistent Deductive System allows drawing useful conclusions from an inconsistent set of formulas.}
\end{itemize}}
\item{\zi{Model Theory Properties}: Presence of a semantic domain and a function mapping expressions to elements of the semantic domain; properties of interest include:
\begin{itemize}
\item{\zi{Truth Valuation System}: The number of, and relationships between truth values (e.g., four truth values with two order relations, as in Belnap's four valued logic \cite{Belnap:1977}).}
\item{\zi{Inconsistency Valuation}: If an expression can be evaluated as both true and false, and what aggregate truth value such expressions obtain.}
\item{\zi{Incompleteness Valuation}: If the truth value of an expression can be undetermined, and what truth value such expressions then obtain.}
\item{\zi{Utility Valuation}: If the truth, falsity, or another valuation of an expression is interpreted as being (and how) valuable with regards to a purpose; e.g., truth of a requirement can be informally interpreted, in an \xc{rem}, as that the requirement will be satisfied by the system-to-be if it is designed according to the Solution of the \xc{rem}, so we see some truth values are being more desirable than others in the resolution of the Requirements Problem.}
\end{itemize}}
\end{itemize}

\subsubsection{Examples}
\xc{kaos}, Formal Tropos and Techne are Multi-Formalism \xc{rem}s, as each includes a Symbolic and a Graphic Syntax: linear-temporal logic and goal trees in \xc{kaos}, linear temporal logic and i* diagrams in Formal Tropos, a custom symbolic syntax with a deductive system and corresponding graphs in Techne. Viewpoints can be a Multi-formalism \xc{rem} if different Viewpoints use different Formalisms. Techne has Syntax Maps which ensure that all represented in Symbolic Syntax can be translated into  Grphical Syntax, and back. This is not the case in \xc{kaos} and Formal Tropos: e.g., temporal relations that can be captured in linear temporal logic have no corresponding representation in Graphical Syntax in either of these \xc{rem}s. Neither \xc{kaos} nor Formal Tropos have a Deductive System, while Techne does. In contrast, Techne has no Model Theory, while both \xc{kaos} and Formal Tropos do. Deductive Systems of both Labeled Quasi-Classical Logic and Techne fail Classicality, both being Paraconsistent; however, they are not Paraconsistent in the same way, meaning that they would not derive the same conclusions given a same set of inconsistent formulas. i* has Graphical Syntax, no Deductive System and no Model Theory. Truth Valuation System in \xc{kaos} and Formal Tropos is that of linear temporal logic, so it can be understood as involving two truth values (if we say ``true'' for a formula which is satisfied, false otherwise), so that there are no interesting Inconsistency and Incompleteness Valuations. Utility Valuation in \xc{kaos} and Formal Tropos is not developed beyond the idea that if a formula representing a requirement is true/satisfied, then this is seen as beneficial.

\subsubsection{Case Study}
Since $\vdash$ in the Requirements Problem is the consequence relation of classical logic, the Formalism of X must be either classical logic, or another formalism which ensures that we can check derivability or satisfaction (e.g., linear temporal logic will work) and has a notion of inconsistency, so that we can check if some given set $R$, $K$, and/or $S$ is consistent (as the formulation $K, S \vdash R$ requires that $K \cup S \cup R$ is consistent). If we add a Graphical Syntax, then X will be a Multi-Formalism. If so, then Syntax Maps define how formulas from the Symbolic Syntax map to (combinations of) primitives in Graphical Syntax. If we keep classical logic as one of the two Formalisms, then it will fail Paraconsistency. The Truth Valuation System properties in that case are also straightforward. Note that there is no Incompleteness and Utility Valuation.



\subsection{Organization Mechanism}\label{s:components:organization-mechanism}

\subsubsection{Definition} Organization Mechanisms are intended to facilitate the creation and manipulation of expressions written in the Formalism of the \xc{rem}. An Organization Mechanism will include rules enabling, e.g., to reuse and combine model fragments, to highlight relations between model fragments that the Formalism cannot or is not intended to show.

\subsubsection{Purposes of Organization Mechanisms in \xc{rem}}
\begin{itemize}
\item{Modularity: A model can be split into pieces and each piece presented individually, perhaps accompanied with comments helping the reader of the model.}
\item{Problem decomposition: Different pieces of the model can focus on different aspects of functionality of the system-to-be. The Organization Mechanism can allow an aspect to be considered while hiding others, and showing only its relationships with others. This can help distribute work among those involved in modeling.}
\item{Reuse: A piece of a model may represent requirements that need to be satisfied by different features of the system-to-be. The Organization Mechanism can allow inclusion of pieces by referencing them, thus avoiding repetition and enabling reuse.}
\end{itemize}

\subsubsection{Component Properties}
\begin{itemize}
\item{\zi{Reference}: Presence of tools to reference, without reproducing, pieces of a model.}
\item{\zi{Structure}: Presence of part-of and is-a relations between model pieces, to capture, respectively, (i) that a piece is an aggregate of other pieces, that the latter are parts of the former, and (ii) that a piece is a generalization of other pieces, i.e., that the latter are specializations of the former.}
\item{\zi{Interface}: Presence of tools to describe how a model piece depends on another model piece, without describing the internals of either, and thus, how model pieces depend on contents of other model pieces, or of operations defined in other model pieces.}
\item{\zi{View}: Presence of tools to group pieces of a model according to interests of different stakeholder groups, such as clients or suppliers, managers or engineers, etc.}
\item{\zi{Constraint}: Ability to define constraints on relationships between pieces of models, such as conditions which should be satisfied for a set of pieces to be parts of another piece, or that some pieces together are a refinement of another piece, etc.}
\item{\zi{Verification}: availability of algorithms to automatically verify whether constraints on relationships between pieces of models are satisfied.}
\end{itemize}

\subsubsection{Examples}
i* allows Structuring via actor boundaries, to indicate that some model pieces belong to the same actor (stakeholder, user, or otherwise). Tropos and Formal Tropos inherit this mechanism. Formal Tropos includes templates, each template being associated to an instance of the ontology in i*. This means that i* models act as-if they are the Organizing Mechanism for a specification in linear temporal logic. In other words, goals, tasks, and other notions in i* models play the same role as schemas and schema relations play in the Z notation: they are used to organize pieces of a model. This idea is in \xc{kaos}, which uses the goal concept and goal trees as an Organization Mechanism for formulas in linear temporal logic. Some relations in \xc{kaos} are defined in a way which allows Verification: e.g., goal refinement is defined via properties between formulas of linear temporal logic in the goals participating in the refinement (consistency of the goals in refinement, minimality of the refining goal set, etc.) so that Verification of such properties is feasible. Labeled Quasi-Classical Logic and Techne have no Organization Mechanism. Viewpoints themselves are an Organization Mechanism of model pieces. \xc{carl} uses its Ontology as an Organizing Mechanism, taking sets of formulas to be extensions of concepts in its Ontology. These examples raise the issue of what interplay there can be between Ontology, Formalism, and Organization Mechanisms in an \xc{rem} (see, \S\ref{s:challenges:integration}).

\subsubsection{Case Study}
For example, to have views in X, we can adopt the ideas from Viewpoints. In this case, we can define meta-level rules for, e.g., solving inconsistencies between viewpoints, including a viewpoint into another viewpoint, and for referencing viewpoints in one another. We can thus use Viewpoints to satisfy Component Properties such as Reference, Structure, View, and Constraint. If meta-level rules are themselves defined in a logic, and there are means for, say, model checking for that logic, X can satisfy Verification.

\subsection{Guidelines}\label{s:components:guidelines}

\subsubsection{Definition} 
Guidelines include all recommendations given on how to instantiate the concepts and relations of the Ontology, make models using the Formalism and manage models using Organization Mechanisms in an \xc{rem}. 

\subsubsection{Purposes of Guidelines in \xc{rem}}
Guidelines suggest how to use the components of an \xc{rem} to accomplish activities in \xc{re}, such as elicitation, modeling, analysis, negotiation, validation, or otherwise.

\subsubsection{Component Properties}
\begin{itemize}
\item{\zi{Design Guidelines}: Presence of rules and steps in which to apply rules to structure the problem space. Design Guidelines are present if it is explained how to refine and operationalize requirements, and identify/define alternative refinements and operationalizations. A refinement relates requirements at different levels of detail; operationalization relates a requirement to resources and processes to use and apply to satisfy the requirement.}
\item{\zi{Decision Making Guidelines}: Presence of rules and steps for defining criteria for ranking alternative refinements and operationalizations of requirements, and select the refinements and operationalizations ranking highest according to most important criteria.}
\item{\zi{Inconsistency Handling Guidelines}: Presence of rules and steps for knowing what model pieces are inconsistent and what to do about them, such as whether to tolerate or resolve the inconsistencies, which of them to resolve when (as soon as detected, or later), etc.}
\item{\zi{Tool Support}: availability of software designed to facilitate the modeling or other applications of the \xc{rem}.}
\end{itemize}

\subsubsection{Examples}
Any \xc{rem} that includes the refinement (or decomposition) and operationalization (or means-ends) relations allows the structuring of the requirements problem and solution space, which includes all \xc{rem}s being able to formalize such relations. This does not mean that they all include Design Guidelines. \xc{kaos}, i*, Tropos, Formal Tropos include guidelines on how to refine/decompose and operationalize requirements. Viewpoints comes with instructions on how to relate and combine viewpoints, and in this sense also provide Design Guidelines. Techne, Labeled Quasi-Classical Logic and \xc{carl} have the necessary concepts and relations, but are not explicit on steps to follow. Decision Making Guidelines are less common. Notions such as quality criteria and nonfunctional requirements are present in Tropos, i*, Techne, but methods on how to rank alternatives are explicit only in \xc{nfr} and in \xc{kaos} in relation to uncertainty \cite{Letier+:2004:FSE}. Labeled Quasi-Classical Logic, \xc{carl} and Viewpoints do not include the necessary concepts, relations, and guidelines. Inconsistency Handling Guidelines are given in Viewpoints, \xc{nfr}, \xc{kaos}, Tropos (as in \xc{nfr}). Techne is paraconsistent, but does not provide explicit Guidelines on what to do, when inconsistency is deduced. \xc{kaos}, i*, \xc{nfr}, Tropos, Viewpoints, \xc{carl} all have software tools to support Guidelines. 

\subsubsection{Case Study}
To make X into a Design Method, we need at least to ensure that it has the refinement relation. To have Decision Making Guidelines in X, we need to add at least one preference relation, rules for aggregating preferences, and a decision rule to rank alternatives. An alternative can amount to a consistent specification $S$, which also satisfies the condition that $K, S \vdash R$. Preference relations would indicate relative desirability of each specification. The rule for aggregating preferences and for ranking alternatives should let us define a total order over all specifications. The idea is then, that we would select the highest-ranking specification. If the Formalism is classical logic and Viewpoints are used as the Organization Mechanism, then Inconsistency Handling Method can be defined using meta-level rules. Finally, Tool Support will be satisfied if there is software that helps make and do reasoning on models made with X.



\section{Research Challenges}\label{s:challenges}
Research challenges can be summarized in the following questions:

\begin{itemize}
\item{How to design \xc{rem} in a systematic way?}
\item{How to validate \xc{rem} in a systematic way?}
\item{How to teach design and validation of \xc{rem}?}
\end{itemize}

These questions become more specific when considered for each Component and Component Property. 

\subsection{What criteria should we use to evaluate the relevance of ontological commitments?}
That is, how to make and justify assumptions and decisions that led to define an \xc{rem}'s Ontology in a particular way, and specifically why the Ontology has the given scope and depth? In the case study, this means explaining why there are three and not more top-level concepts in X, why some of the three are not specializations of others, whether some or all of top level concepts should be specialized in the Ontology, and if yes, then to what depth. 

There are different complementary methods for answering these questions. Justification for the three concepts can be given in terms of arguments against ontologies in existing \xc{rem}, by deriving the Ontology from a more general body of knowledge (e.g., claiming that concepts should cover specific grammatical moods), deriving the Ontology from a higher-level (e.g., a foundational) ontology, or justifying concepts by the presence/absence of some specific information in many experience reports and case studies. Relative merits and limitations of these different approaches are not clear enough, making it difficult to say how one should approach Ontology engineering for an \xc{rem}.

\subsection{How can we inform the design of Ontology and Formalism, through empirical research into categories of information and reasoning rules that engineers tend to recurrently use or disregard during \xc{re}?}\label{s:challenges:validation}
Leaving aside the application of \xc{rem} to case studies as a form of validation, empirical research can be done to inform Ontology and Formalism design for \xc{rem}. Empirical research on human nonmonotonic reasoning suggests a direction. Namely, just as factors influencing human nonmonotonic reasoning have been studied (cf., e.g., \cite{Ford+2000:CI}), so can factors influencing reasoning about requirements problems be studied. In the former, data collection consists of asking subjects to choose among predefined answers to problems requiring nonmonotonic reasoning as formalized in, say, default logic. If such an approach is applied to evaluate \xc{rem}s, then observing systematic departure in answers people give to a specific problem of modeling or reasoning about requirements, from answers that an \xc{rem} would provide, can suggest that \xc{rem} helps reduce error in that specific modeling and/or reasoning task.


\subsection{Can there be a core ontology applicable across \xc{rem}s?} An ontology is a core ontology if it is minimal with regards to a purpose, i.e., includes only non-overlapping concepts and relations that are necessary and sufficient for satisfying a purpose. In the case study, the minimal ontology includes the Requirement, Specification, and Domain assumption concepts, along with only those relations necessary and sufficient to capture the complex relationship $K, S \vdash R$ (complex, because to define $\vdash$, one uses connectives over formulas in $K$, $S$, $R$ and relations between premises and conclusions in proof rules of classical logic). To have a core ontology applicable across \xc{rem}s requires recognizing and successfully arguing that there are concepts and relations without which a proposal for an \xc{rem} fails. This then leads to questions such as, Can there be an \xc{rem} which has Design Guidelines, but which cannot model the refinement relation?, or Can an \xc{rem} support decision making by modeling alternative solutions to a requirements problem, while not having some form of the preference relation between requirements? So if one claims an \xc{rem} must have Design Guidelines (for how would it otherwise solve a requirements problem which assumes unclear and incomplete requirements are what we start with in \xc{re}?), then one needs also to choose whether refinement is a core relation. Not only this, but one also has to determine if there are relations from which refinement can be defined (e.g., as in Techne), so that refinement, while perhaps necessary in any \xc{rem}, really is a derived relation, not a primitive one.

\subsection{Can we further clarify the role that concepts and relations have in reasoning about requirements?}\label{s:challenges:meaning-is-use}
There is an important difference between \xc{rem}s such as \xc{kaos}, Tropos, Techne, i* and Formal Methods such as Z and Larch. In Z, the concept of \zi{Schema} groups definitions and expressions. In the mentioned \xc{rem}s, the Goal concept (and other concepts) are intended to do more than organize formulas. The Goal concept is particularly illustrative, in that it says the conditions stated in the formulas (the linear temporal logic formulas in \xc{kaos} and Formal Tropos) it ``includes'' \zi{are desired}, which is something that these formulas alone do not convey (as there is no sort, modality, or otherwise in linear temporal logic which refers to desirability). \zi{But} these same \xc{rem}s do not take the next step, one analogous to what modal logics do with regards to classical logic: the \xc{rem}s do not study how to introduce these modalities into the semantics of the Formalism in the \xc{rem}. The very specific question is, for example, does the Goal concept give a sort on formulas in a Formalism, and if yes, then does this sort merely label formulas, or does the sort of the formula influence the role this formula has in proof theory of that Formalism? 

To make the point here clearer, consider the case study again. If we take classical logic, and make its language sorted, with the three sorts $R$, $K$, and $S$, but at the same time, we keep the semantics of classical logic, we have only introduced labels on formulas: reasoning is still that of classical logic, and so we have failed to capture, \zi{in the Formalism}, the intuitive ideas on what it means for a formula to be a Requirement, while another one is a Domain assumption. 

If we follow Wittgenstein's aphorism that ``meaning is use'', then \xc{rem}s such as \xc{kaos}, Tropos, Formal Tropos, Techne, along with all those cited in this paper are limited, since they use pre-existing logics which disregard these sorts defined by the Ontology of the \xc{rem}. 

\zi{If we wanted conceptual tools designed to the specific purpose of the \xc{rem}}, then our intuitions about the difference \zi{in use and during} \xc{re} of $R$, $K$, and $S$ expressions, should be reflected in the rules used to draw conclusions \zi{in the Formalism}. So if we focus on proof theory, we would want to ``embed'' whatever meaning we have in mind for $R$, $K$, and $S$ \zi{by the proof theory itself}, precisely in order to make sure that the conclusions are drawn in a way that satisfies the intended meaning. 

This brings me back to the question of how we could further clarify the role of concepts and relations in reasoning about requirements. We can do this -- namely, embed the intended interpretation of concepts and relation of an \xc{rem}'s Ontology into its Formalism -- \zi{only if we have clarified the role that these concepts and relations have when reasoning about requirements in that \xc{rem}}, i.e., only if we are very clear on their use, and from there, of their meaning. 

To put it plainly: it is relatively easy to say that some formula $\phi$ is an instance of the Requirement concept; what is harder is explaining how this additional information -- that $\phi$ is a Requirement, not a Domain assumption -- influences the conclusions we will draw about the satisfaction of $\phi$, or about inconsistency between $\phi$ and some other formulas. 

To illustrate this, suppose that we want to have a knowledge base which includes all formulas with K, S, and R labels. If that knowledge base makes deduction using the classical $\vdash$, then the question ``Is a requirement $\phi$ satisfied?'' gets an irrelevant answer in at least two cases:
\begin{enumerate}
\item{If the knowledge base is inconsistent, then $\vdash$ will derive $\phi$ regardless of what actually is in that knowledge base (just as it will derive \zi{any} other formula, because of \zi{ex falso quodlibet}).}
\item{Because $\vdash$ is reflexive, meaning that any formula on its left-hand side is always deduced, we will conclude that any formula on the left-hand side is satisfied (formally, for any set $X$ of formulas, and any formula $\phi$, we have that $X \cup \{ \phi \} \vdash \phi$).}
\end{enumerate}

Yet in both cases, it makes no sense to say that $\phi$ is satisfied, for the simple reason that in both cases, we have said nothing about if $\phi$ is operationalized, refined, or otherwise.

Now, observe that the knowledge base will give wrong answers \zi{not} because $\vdash$ is somehow deficient by itself, \zi{but because the proof theory defining $\vdash$ sees no difference between formulas that are requirements, domain assumptions, and specifications (or any other category one deems relevant)}. We can respond to this in two ways:
\begin{enumerate}
\item{We can make tools that are \zi{outside} the knowledge base, and which filter, \zi{after} deduction, the results of deduction by applying some rules. This is what happens in \xc{kaos} for example (although it is not about deduction, but model checking, but that makes no difference here), as it requires first that conflicts and obstacles be eliminated to repair consistency, and only then can questions, such as whether a requirement is satisfied, be asked.}
\item{We can make a Formalism which is attentive to which formulas instantiate which concepts and relations from the Ontology. But this requires a considerable change in how \xc{rem}s are made, as the following question suggests.}
\end{enumerate}

\subsection{Integration of Components}\label{s:challenges:integration}
It is without doubt good for a representation of requirements to be modular. But it is not clear whether it is good for an \xc{rem} to be modular. All \xc{rem}s mentioned in this paper are modular in the following sense: \zi{\xc{rem} Components are designed to a considerable extent independently from the Formalism component}. This is a strong claim, but one not difficult to argue for. 

Take Tropos as an example. It uses the Ontology of i*. One of its two Formalisms is the graphical language of i*. The other is linear temporal logic. The language of i* is obviously defined to fit the Ontology of i*. But linear temporal logic is independent from the i* Ontology and from the i* Formalism. Yet it is via linear temporal logic that one can do automated reasoning in Tropos. Just as linear temporal logic ignores that there are actors, goals, tasks in i* models, so it ignores that there are goals, agents, refinements in \xc{kaos} models. 

For further illustration, take the \xc{rem} X in the case study. As mentioned earlier (cf., \S\ref{s:challenges:meaning-is-use}), if X has classical logic as its Formalism, then asking questions to the knowledge base of X will give misleading answers. That is, the answers are likely to violate the rules and processes stated in the Guidelines of the \xc{rem}. One such rule is that a requirement must be satisfied by functionality described in the specification $S$ and which is consistent with the domain assumptions $K$. Yet, we would still get the answer from the knowledge base that a requirement $\phi$ is satisfied, even when $S$ is inconsistent and includes no descriptions of functionality for performing tasks which satisfy $\phi$.

If one prefers to think in terms of goals and tasks that satisfy the goals, then suppose $\phi$ is a goal (i.e., $\phi$ is an instance of the \zi{Goal} concept in the Ontology of the \xc{rem}). Suppose that the guideline in the \xc{rem} is this: a goal is satisfied if there are tasks that operationalize it, meaning that if these tasks are satisfied, then the goal is satisfied as well. If \xc{rem} uses classical logic as its Formalism, then deducing $\phi$ does not mean it is satisfied in the said sense, because (as mentioned above -- cf., \S\ref{s:challenges:meaning-is-use}) it can be deduced in cases when there are no tasks which operationalize $\phi$.

\section{Discussion}\label{s:discussion}
The Research Challenges section (cf., \S\ref{s:challenges}) started by asking three questions, namely, how to (i) systematically design \xc{rem}, (ii) validate them, and (iii) teach design and validation. 

It should be clear that the proposed classification framework has a limited use in answering these questions. For systematic design, the framework gives a checklist of ingredients of \xc{rem} that a \xc{rem} designer will in one way or another need to think about. This checklist itself suggests what knowledge one will need to apply when designing an \xc{rem} -- ontology engineering, formal logic, process design, etc., \zi{in addition to} her understanding of \xc{re}. The framework does not say what concepts and relations are more relevant than others, what reasoning rules to use, etc. For validation, the framework suggests how validation methods already known for specific Components and Component Properties can be used for validation of an \xc{rem} (cf., \S\ref{s:challenges:validation}). For teaching, the framework suggests the topics to cover with students and researchers interested in the application, engineering, extension, and validation of \xc{rem}s. 

The rest of this section discusses the relationship between the concept of \xc{rem} and concepts of Requirements Modeling Language and Formal Method.

\subsection{Requirements Modeling Language}
The emphasis in a Requirements Modeling Language is on \zi{language}, i.e., a conceptual tool for representation of, and reasoning about requirements. In an \xc{rem}, this conceptual tool would amount to the combination of Ontology, Formalism, and Organization Mechanism components. In Requirements Modeling Languages such as \xc{rml} and i*, guidelines for the use of the language are usually treated separately, as are the Requirements Problem and Solution concepts. An \xc{rem} can be viewed as including an \xc{rml}, if we take an \xc{rml} to include the Ontology, Formalism, and Organization Mechanism components.

\subsection{Formal Methods}
I take Wing's definition of Formal Methods \cite{Wing:1990:C} (\xc{fm}s hereafter) as the definition of \xc{fm}s. 

In terms of Components, \xc{fm}s are combinations of Formalism, Organization Mechanisms, and Guidelines. Ontology is not developed in the same sense as in \xc{rem}s, and at best can amount to syntactic sugar, to make specifications readable by customers, in addition to specifiers and implementors. The classification framework suggests that Requirements Problem and Solution, Ontology, and Guidelines in an \xc{rem} are not merely syntactic sugar, but, by influencing the conceptualization of the requirements problem and the process of its resolution, influence how one designs, or should design, the Formalism and Organization Mechanisms. 

We can, so to speak, hack an \xc{fm} by adding an Ontology to it as syntactic sugar, and so make it look like an \xc{rem}. The limitation of doing so is that the hacked artifact -- the \xc{fm} -- shows its limitations as soon as we start using it to \xc{re}-specific tasks, which were not of interest to the designer of the \xc{fm}. An example is to make knowledge bases using the Formalism in the \xc{rem}, and to ask questions about which requirements are satisfied. If Ontology is only syntactic sugar and Guidelines are merely text alongside the \xc{fm}, we are likely to get wrong answers (for reasons stated earlier -- cf., \S\ref{s:challenges:meaning-is-use}). This is not the problem of the underlying \xc{fm}, but of the fact we are using it despite knowing that it ignores Ontology added on top of it and the Guidelines for its use, so that it cannot make sure its answers reflect the knowledge that the Ontology and Guidelines capture. In a summary, \xc{rem}s are \zi{not} specializations of \xc{fm}s.

\section{Conclusions}\label{s:conclusions}
This paper suggests a classification framework for Requirements Engineering Methods (\xc{rem}s). The framework categorizes \xc{rem}s by the properties of \xc{rem} components. The framework is intended to help the analysis, teaching, and extension of existing \xc{rem}s, and the engineering and validation of new \xc{rem}s. The paper discusses research challenges highlighted by the framework. The framework clarifies the relations between the concept of \xc{rem} and other concepts of interest in and to \xc{re}, and in particular, Requirements Problem and Solution, Requirements Modeling Language, and Formal Method. As noted in the Introduction, this classification framework focuses on one dimension only -- the components of \xc{rem} -- while other dimensions of classification are not discussed here.

The classification framework identifies, through Components and Component Properties, the knowledge applicable when designing \xc{rem}s. In doing so, the framework suggests fragments for a body of knowledge of a research methodology proper to the design of \xc{rem}s. To the extent that \xc{rem} design and validation are important activities in \xc{re} research, the framework contributes to forming a research methodology specific to \xc{re}.

\subsection*{Acknowledgments}

Since 2008, I had discussed the ideas behind this classification framework with many colleagues at University of Trento, University of Toronto, Fondazione Bruno Kessler, and University of Namur. I am indebted to John Mylopoulos and Alexander Borgida for introducing me to requirements modeling languages. I thank St\'{e}phane Faulkner, Neil Ernst, Sotirios Liaskos, Alexei Lapouchnian, Angelo Susi, and Anna Perini for discussions on topics related to \xc{rem}s. This does not mean that they agree with me on this classification framework.

\bibliographystyle{plain}
\bibliography{techne}

\end{document}